\documentclass[a4paper,11pt]{article}
\usepackage[pdftex]{graphicx}
\pdfoutput=1
\usepackage{amsmath}
\usepackage{amssymb}
\usepackage{accents}
\usepackage{color,calc}
\usepackage{wrapfig}
\usepackage{amsthm}

\newcommand{\commentout}[1]{}

\newcommand{\onlystats}[1]{#1}
\newcommand{\toomuchstats}[1]{}
\newcommand{\fabulous}[1]{}

\onlystats{

}

\newcommand{\showguide}[1]{{\em #1}}
\newcommand{\shortrefs}[1]{#1}

\renewcommand{\showguide}[1]{}
\renewcommand{\shortrefs}[1]{}

\definecolor{shade}{gray}{0.8}

\newcommand{\axiom}[1]{{\bf A#1}}
\newcommand{\corr}[1]{{\bf C#1}}
\newcommand{\man}{{\bf M}}
\newcommand{\god}{{\bf G}}
\newcommand{\bad}{{\bf B}}
\newcommand{\good}{{\bf $\neg$B}}
\newcommand{\exist}{{\bf E}}
\newcommand{\notexist}{{\bf $\neg$E}}

\title{God exists with probability $\frac{1}{H+1}$}
\author{Jesse Hoey}
\begin{document}
\maketitle

\begin{abstract}
This note will address the issue of the existence of God from a game theoretic perspective. We will show that, under certain assumptions, man~\footnote{ we use the term {\em man} to denote either gender, as in {\em hu-man}.} cannot simultaneously be (i) rational and (ii) believe that an infinitely powerful God exists.   Game theory and decision theory have long been used to address this thorny question~\cite{PascalPensees1669}.
\end{abstract}

\section{Introduction}
We start with the following assumptions, where we use \man{} to denote Man and \god{} to denote God:
\begin{description}
\item[A1]: \god{} is rational 
\item[A2]: \man{}  is rational
\item[A3]: \man{} has a belief over how bad \god{}'s punishment or (or how good \god{}{}'s reward) can be in the afterlife, $H$, in units of how bad (or good) life on earth is. This distribution will be denoted $P_m(H)$, and that this distribution is bounded bewteen $0$ and some $H_{max}$, such that $P_m(h)=0$ if $h<0$ or if $h>H_{max}$.  Note that $H$ may be a discounted estimate (so paying a cost of $h$ in 20 years is preferable to paying this cost now).
\item[A4]: \man{} is not perfect (is not perfectly bad or good)
\end{description}

Assumption \axiom{1} and \axiom{2} are both debatable. Assumption \axiom{3} relies on \man{}'s inability to comprehend infinity, and to form a belief distribution over unknown future rewards. The assumption of a maximum penalty (reward) $H_{max}$ may be removable.   Assumption \axiom{4} is the easiest to accept, since if A4 were not true, then \man{}$\equiv$\god{}. 

\begin{figure}[h]
\begin{center}
\begin{tabular}{ll|cc|}
& & \multicolumn{2}{c}{\god{}}\\
& & \exist{}  & \notexist{} \\  \hline
\man{} & \good{} & $(+H,+X)$ & $(-1,0)$ \\
& \bad{} & $(-H,-X)$ & $(+1,0)$ \\ \hline
\end{tabular}
\end{center}
\caption{\label{payoffmat} Payoff matrix for the \man{}-\god{} game.}
\end{figure}
 Now, we make the following simplifying assumptions.
% (which are not strictly necessary...still need to show this). 
 \man{} can either be good (\good{}) or bad (\bad{}), and \god{} can either exist (\exist{}) or not exist (\notexist{}).  \god{} gets a payoff of $0$ if he doesn't exist, and a payoff of $+X$ if he does exist and \man{} is good,  and a payoff of $-X$ if he does exist and \man{} is bad.  Here we are assuming \god{} has some incentive for trying to get \man{} to be \good{}.  If \god{} doesn't exist, then \man{} gets a payoff of $+1$ if he is \bad{} and $-1$ if he is \good{} (so being bad is fun, so long as there are no consequences).  If \god{} does exist, then \man{} gets a payoff of $+H$ if he is \good{} (so he goes to heaven), and $-H$ if he is \bad{} (so he goes to hell).  Note that, at this stage, we are only assuming $H>0$ without loss of generality. We are also, wlg, making \man{}'s payoffs if \god{} exists in units of his payoffs when \god{} doesn't exist.   The fact that we are using $\pm H$ is also wlg.    In this case, the payoff matrix is as shown in Figure~\ref{payoffmat}.

There are three equilibria of this game, two pure ones at ([\good{}],[\exist{}]) and ([\bad{}],[\notexist{}]) and one randomized one ($\frac{1}{2}$[\good{}]+$\frac{1}{2}$[\bad{}],$\frac{1}{H+1}$[\exist{}]+$\frac{H}{H+1}$[\notexist{}]).  See Section~\ref{sec:equils} for a derivation of these formulae.

Now, we can draw the following intermediate corollaries
\begin{description}
\item[C1]: From \axiom{4} we can deduce that \man{} must play the randomized strategy (since \man{} cannot play either pure strategy [\good{}] or [\bad{}]
\item[C2]: From \axiom{1}$\wedge$\axiom{2}$\wedge$\corr{1} we can deduce that \god{} must also play a randomized strategy.  This is because if \god{} did not do so, then \man{} would not play a randomized strategy ($\neg$\corr{1}), but \man{} does (\corr{1}), a contradiction. 
\item[C3]: From \corr{1}$\wedge$\corr{2} and the equilibria of the game, we can deduce that \man{} must play $\frac{1}{2}$[\good{}]+$\frac{1}{2}$[\bad{}] (is good $50\%$ of the time) and \god{} must play $\frac{1}{H+1}$[\exist{}]+$\frac{H}{H+1}$[\notexist{}] (exists with probability $\frac{1}{H+1}$)
\end{description}

\begin{figure}[h]
\begin{center}
\includegraphics[width=0.4\textwidth]{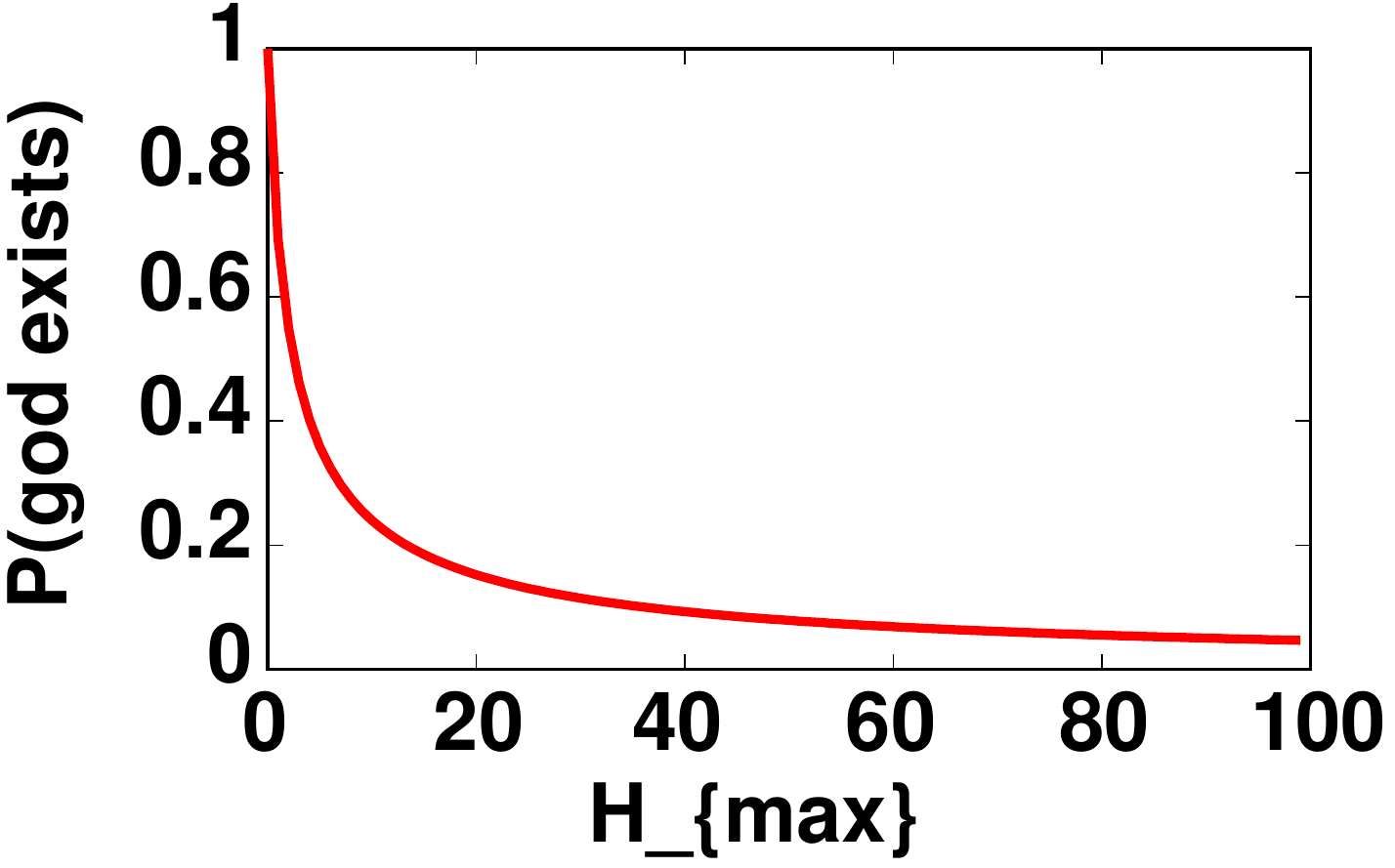}
\end{center}
\caption{\label{godp} Probability that \god{} exists as a function of $H_{max}$}
\end{figure}

Therefore, we have shown that \god{} exists with probability $\frac{1}{H+1}$.  However, what is $H$? In \axiom{3}, we assumed that \man{} has a belief distribution over $H$, $P_m(H)$.  Therefore, we can deduce that
\[ P(\exist{}) = E_{P_m}\left[ \frac{1}{H+1} \right]= \int_{-\infty}^{+\infty} \frac{1}{H+1}P_m(H)dH\]
Let's suppose that $P_m(H)$ is a constant over the interval $0<H<H_{max}$, then 
\[P(\exist{}|H_{max}) = \int_{0}^{H_{max}} \frac{1}{H+1}\frac{1}{H_{max}}dH = \frac{\log(H_{max}+1)}{H_{max}}\]
Figure~\ref{godp} shows the probability that \god{} exists as a function of $H_{max}$
We can deduce certain interesting properties of this existence probability.   They center around the limits of $H_{max}$, and we denote $H_{max}\rightarrow\infty$ (an infinitely powerful \god{}) by $G_{\infty}$, and $H_{max}\rightarrow 0$ (a powerless \god{}) by $G_0$.
\begin{description}
\item[T1]: $G_{\infty}\rightarrow P(\exist{})=0$.  Thus, if one believes that \god{} is infinitely powerful,  \god{} does not exist. 
\item[T2]: $G_0\rightarrow P(\exist{})=1$. Thus, if one believes that \god{} is powerless, \god{} exists. 
\item[T3]: if \man{} and \god{} are rational, and \man{} is imperfect, then \man{} cannot hold that $G_{\infty}$.
\end{description}
Thus, if \man{} and \god{} are rational, and \man{} is imperfect, then \god{} either does not exist, or is not infinitely powerful.   This holds under a wide range of $P_m(H)$, but exactly how wide this range is we leave as an exercise for the interested reader. 

So, the bottom line is that, if  Man believes that God exists, then sin is irrational.  Therefore, since Man is not perfect, then belief in god implies irrationality. Man is thus put in a difficult position, but cannot have his cake and eat it too: he cannot be rational and believe in God unless he can be perfect.  But if he were perfect, then he would be God, meaning only that he could be rational only if he were God.  Being not-God, he must therefore be irrational if he wants to believe in God.  Therefore, rational man is forced to (1) not be an atheist, and (2) not believe that God exists with probability 1. Instead, he must have a belief distribution over the probability that God exists, and play his randomized strategy of being bad half the time.  

\section{Derivation of Equilibria}
\label{sec:equils}
At these equilibria, neither player has any incentive to change their strategy.  Consider the pure strategies first.  If \exist{}, then \man{} will have no reason to be bad (\bad{}), as his payoff would be $-H$ instead of $+H$ for being good.  At the same time, \god{} has no reason to not exist, as he will lose the $+1$ he gets if he does exist and \man{} is \good{}.  A similar argument holds for the second pure strategy equilibrium.  The randomized equilibrium is more interesting, and can be computed as follows.   \man{} would be willing to randomize (play $p_m$[\good{}]+$(1-p_m)$[\bad{}]) if playing \good{} or \bad{} would give him the same payoffs, so we can ask with what probability, $p_g$, would \god{} need to randomize (play $p_g$[\exist{}]+$(1-p_g)$[\notexist{}]) in order for this to happen.  This will be the point at which \man{}'s payoff for playing \good{} are the same as for playing \bad{}:
\[ p_g\times H+(1-p_g)\times(-1) = p_g\times(-H)+(1-p_g)\times 1\]
solving this yields $p_g=\frac{1}{H+1}$.  This means that \man{} would be willing to play his randomized strategy if \god{} played a randomized strategy of $\frac{1}{H+1}$[\exist{}]+$(1-\frac{1}{H+1})$[\notexist{}].  What is \man{}'s strategy though?  We can apply the same analysis, and consider that \god{} will only randomize if each of his possible actions (\exist{} and \notexist{}) will pay off the same, which is when
\[ p_m\times 1 + (1-p_m)\times(-1) = p_m\times 0 + (1-p_m)\times 0\]
solving this yields $p_m=\frac{1}{2}$.  Thus, \god{} would be willing to play (any) randomized strategy if \man{} played $\frac{1}{2}$[\good{}]+$\frac{1}{2}$[\bad{}], and \man{} would only play such a strategy if \god{} played exactly the randomized strategy $\frac{1}{H+1}$[\exist{}]+$(1-\frac{1}{H+1})$[\notexist{}].   

\section{Related Work}
Pascal was the first to write down a game-theoretic analysis of God's existence in his ``Wager''~\cite{PascalPensees1669}.  He is probably using the payoff matrix show in Figure~\ref{payoffmat-pascal}.  He uses the notion of ``wager'' for/against god in place of our \good{}/\bad{}, but the interpretation is the same. 
\begin{figure}[h]
\begin{center}
\begin{tabular}{ll|cc|}
& & \multicolumn{2}{c}{\god{}}\\
& & \exist{}  & \notexist{} \\  \hline
\man{} & \good{} & $(+H,?)$ & $(0,?)$ \\
& \bad{} & $(-H,?)$ & $(0, ?)$  \\ \hline
\end{tabular}
\end{center}
\caption{\label{payoffmat-pascal} Payoff matrix for the \man{}-\god{} game (from~\cite{sep-pascal-wager})}
\end{figure}

In this case, there is only one pure-strategy equilibrium, at ([\good{}],[\exist{}]), and Pascal uses this a proof that \god{} exists.  However, he does not show any payoffs for \god{}, apparently assuming that all payoffs are equal for\god{}, and thus that he is playing a randomized strategy of ($\frac{1}{2}$[\exist{}],  $\frac{1}{2}$[\notexist{}]).  Note that, if this were the case, then \man{} would still only play \good{}.  However, Pascal also notes there are problems with this matrix, in fact that assigning the same outcome ($0$) to \man{} if \notexist{} is flawed (reproduced from~\cite{sep-pascal-wager}):
\begin{quote}
{\em ``The thought seems to be that if I wager for God, and God does not exist, then I really do lose something. In fact, Pascal himself speaks of staking something when one wagers for God, which presumably one loses if God does not exist. (We have already mentioned ‘the true’ as one such thing; Pascal also seems to regard one's worldly life as another.) In other words, the matrix is mistaken in presenting the two outcomes under ‘God does not exist’ as if they were the same, and we do not have a case of superdominance after all.''}
 \end{quote}

 Pascal further asserts that (reproduced from~\cite{sep-pascal-wager}):
\begin{quote}
{\em ``As for the utilities of the outcomes associated with God's non-existence, Pascal tells us that “what you stake is finite”. This suggests that whatever these values are, they' are finite.''}
\end{quote}
Thus, this seems to indicate that \man{}'s payoffs when \notexist{} should be some finite penalty for \good{}, as we have done in Figure~\ref{payoffmat}.

\bibliographystyle{plain}
\bibliography{../../refs}

\begin{thebibliography}{1}

\bibitem{sep-pascal-wager}
Alan Hájek.
\newblock Pascal's wager.
\newblock In Edward~N. Zalta, editor, {\em The Stanford Encyclopedia of
  Philosophy}. Summer 2011 edition, 2011.

\bibitem{PascalPensees1669}
Blaise Pascal.
\newblock {\em Pens\'{e}es}.
\newblock Guillaume Desprez, Paris, 1669.

\end{thebibliography}

\end{document}